\documentstyle[12pt]{article}
\begin{document}

\def\RN{Reisner-Nordstr\"om }
\def\Sch{Schwarzchild }
\def\HH{Hartle-Hawking }
\def\be{\begin{equation}} \def\bea{\begin{eqnarray}}
\def\ee{\end{equation}}\def\eea{\end{eqnarray}}
\def\ncr{\nonumber\\ }
\def\MBVR{Maja Buri\'c
\footnote{E-mail: majab@rudjer.ff.bg.ac.yu} and
 Voja Radovanovi\'c\footnote{E-mail: rvoja@rudjer.ff.bg.ac.yu}\\
{\it Faculty of Physics, P.O. Box 368, 11001 Belgrade, Yugoslavia}}

 
\let\und=\b                     
\let\ced=\c                     
\let\du=\d                      
\let\um=\H                      
\let\sll=\l                     
\let\Sll=\L                     
\let\slo=\o                     
\let\Slo=\O                     
\let\tie=\t                     
\let\br=\u                      

 
\def\a{\alpha}
\def\b{\beta}
\def\c{\chi}
\def\d{\delta}
\def\e{\epsilon}                
\def\f{\phi}                    
\def\g{\gamma}
\def\h{\eta}
\def\i{\iota}
\def\j{\psi}
\def\k{\kappa}
\def\l{\lambda}
\def\m{\mu}
\def\n{\nu}
\def\o{\omega}
\def\p{\pi}                     
\def\q{\theta}                  
\def\r{\rho}                    
\def\s{\sigma}                  
\def\t{\tau}
\def\u{\upsilon}
\def\x{\xi}
\def\z{\zeta}
\def\D{\Delta}
\def\F{\Phi}
\def\G{\Gamma}
\def\J{\Psi}
\def\L{\Lambda}
\def\O{\Omega}
\def\P{\Pi}
\def\Q{\Theta}
\def\S{\Sigma}
\def\U{\Upsilon}
\def\X{\Xi}
 
 
\def\ca{{\cal A}}
\def\cb{{\cal B}}
\def\cc{{\cal C}}
\def\cd{{\cal D}}
\def\ce{{\cal E}}
\def\cf{{\cal F}}
\def\cg{{\cal G}}
\def\ch{{\cal H}}
\def\ci{{\cal I}}
\def\cj{{\cal J}}
\def\ck{{\cal K}}
\def\cl{{\cal L}}
\def\cm{{\cal M}}
\def\cn{{\cal N}}
\def\co{{\cal O}}
\def\cp{{\cal P}}
\def\cq{{\cal Q}}
\def\car{{\cal R}}
\def\cs{{\cal S}}
\def\ct{{\cal T}}
\def\cu{{\cal U}}
\def\cv{{\cal V}}
\def\cw{{\cal W}}
\def\cx{{\cal X}}
\def\cy{{\cal Y}}
\def\cz{{\cal Z}}

 
\def\bo{{\raise.05ex\hbox{\large$\Box$}\:}}             
\def\cbo{{\,\raise-.15ex\Sc [\,}}                       
\def\pa{\partial}                                       
\def\de{\nabla}                                         
\def\dell{\bigtriangledown}                             
\def\su{\sum}                                           
\def\pr{\prod}                                          
\def\iff{\leftrightarrow}                               
\def\conj{{\hbox{\large *}}}                            
\def\ltap{\raisebox{-.4ex}{\rlap{$\sim$}} \raisebox{.4ex}{$<$}}   
\def\gtap{\raisebox{-.4ex}{\rlap{$\sim$}} \raisebox{.4ex}{$>$}}   
\def\TH{{\raise.2ex\hbox{$\displaystyle \bigodot$}\mskip-4.7mu \llap H \;}}
\def\face{\hbox{\normalsize$\;\;\:{\raise.9ex\hbox{\oo n}\mskip-13mu \llap
        {${\buildrel{\hbox{\frtnrm ..}}\over\smile}$}}\:$}}     
\def\Face{{\raise.2ex\hbox{$\displaystyle \bigodot$}\mskip-2.2mu \llap {$\ddot
        \smile$}}}                                      
\def\dg{\sp\dagger}                                     
\def\ddg{\sp\ddagger}                                   
\def\Lhat{{\bf\rlap{\kern-.09em$\hat{\phantom L}$}L}}
\def\Lcheck{{\bf\rlap{\kern-.09em$\check{\phantom L}$}L}}
 
 
\def\sp#1{{}^{#1}}                              
\def\sb#1{{}_{#1}}                              
\def\oldsl#1{\rlap/#1}                          
\def\sl#1{\rlap{\hbox{$\mskip 1 mu /$}}#1}      
\def\Sl#1{\rlap{\hbox{$\mskip 3 mu /$}}#1}      
\def\SL#1{\rlap{\hbox{$\mskip 4.5 mu /$}}#1}    
\def\Tilde#1{\widetilde{#1}}                    
\def\Hat#1{\widehat{#1}}                        
\def\Bar#1{\overline{#1}}                       
\def\bra#1{\Big\langle #1\Big|}                 
\def\ket#1{\Big| #1\Big\rangle}                 
\def\VEV#1{\Big\langle #1\Big\rangle}           
\def\brak#1#2{\Big\langle #1\Big|#2\Big\rangle}         
\def\abs#1{\Big| #1\Big|}                       
\def\sbra#1{\left\langle #1\right|}             
\def\sket#1{\left| #1\right\rangle}             
\def\svev#1{\left\langle #1\right\rangle}       
\def\sabs#1{\left| #1\right|}                   

\def\leftrightarrowfill{$\mathsurround=0pt \mathord\leftarrow \mkern-6mu
        \cleaders\hbox{$\mkern-2mu \mathord- \mkern-2mu$}\hfill
        \mkern-6mu \mathord\rightarrow$}
\def\dvec#1{\vbox{\ialign{##\crcr
        \leftrightarrowfill\crcr\noalign{\kern-1pt\nointerlineskip}
        $\hfil\displaystyle{#1}\hfil$\crcr}}}           
\def\dt#1{{\buildrel {\hbox{\LARGE .}} \over {#1}}}     
\def\dtt#1{{\buildrel \bullet \over {#1}}}              
\def\ddt#1{{\buildrel {\hbox{\LARGE .\kern-2pt.}} \over {#1}}}
\def\der#1{{\pa \over \pa {#1}}}                
\def\fder#1{{\d \over \d {#1}}}                 
 
 
\def\frac#1#2{{\textstyle{#1\over\vphantom2\smash{\raise.20ex
        \hbox{$\scriptstyle{#2}$}}}}}                   
\def\ha{\frac12}                                        
\def\sfrac#1#2{{\vphantom1\smash{\lower.5ex\hbox{\small$#1$}}\over
        \vphantom1\smash{\raise.4ex\hbox{\small$#2$}}}} 
\def\bfrac#1#2{{\vphantom1\smash{\lower.5ex\hbox{$#1$}}\over
        \vphantom1\smash{\raise.3ex\hbox{$#2$}}}}       
\def\afrac#1#2{{\vphantom1\smash{\lower.5ex\hbox{$#1$}}\over#2}}    
\def\tder#1#2{{d #1 \over d #2 }}                 
\def\partder#1#2{{\partial #1\over\partial #2}}   
\def\brkt#1#2{{\left\langle #1 | #2 \right\rangle}} 
\def\secder#1#2#3{{\partial~2 #1\over\partial #2 \partial #3}}  
\def\on#1#2{\mathop{\null#2}\limits~{#1}}       
\def\On#1#2{{\buildrel{#1}\over{#2}}}           
\def\under#1#2{\mathop{\null#2}\limits_{#1}}    
\def\bvec#1{\on\leftarrow{#1}}                  
\def\oover#1{\on\circ{#1}}                              
 
 
\def\boxes#1{
        \newcount\num
        \num=1
        \newdimen\downsy
        \downsy=-1.64ex
        \mskip-7.8mu
        \bo
        \loop
        \ifnum\num<#1
        \llap{\raise\num\downsy\hbox{$\bo$}}
        \advance\num by1
        \repeat}
\def\boxup#1#2{\newcount\numup
        \numup=#1
        \advance\numup by-1
        \newdimen\upsy
        \upsy=.82ex
        \mskip7.8mu
        \raise\numup\upsy\hbox{$#2$}}
 
 
\newskip\humongous \humongous=0pt plus 1000pt minus 1000pt
\def\caja{\mathsurround=0pt}
\def\eqalign#1{\,\vcenter{\openup2\jot \caja
        \ialign{\strut \hfil$\displaystyle{##}$&$
        \displaystyle{{}##}$\hfil\crcr#1\crcr}}\,}
\newif\ifdtup
\def\panorama{\global\dtuptrue \openup2\jot \caja
        \everycr{\noalign{\ifdtup \global\dtupfalse
        \vskip-\lineskiplimit \vskip\normallineskiplimit
        \else \penalty\interdisplaylinepenalty \fi}}}
\def\li#1{\panorama \tabskip=\humongous                         
        \halign to\displaywidth{\hfil$\displaystyle{##}$
        \tabskip=0pt&$\displaystyle{{}##}$\hfil
        \tabskip=\humongous&\llap{$##$}\tabskip=0pt
        \crcr#1\crcr}}
\def\eqalignnotwo#1{\panorama \tabskip=\humongous
        \halign to\displaywidth{\hfil$\displaystyle{##}$
        \tabskip=0pt&$\displaystyle{{}##}$
        \tabskip=0pt&$\displaystyle{{}##}$\hfil
        \tabskip=\humongous&\llap{$##$}\tabskip=0pt
        \crcr#1\crcr}}
 
 
\def\phil{@{\extracolsep{\fill}}}
\def\unphil{@{\extracolsep{\tabcolsep}}}
 

\def\CMP{Commun. Math. Phys.}
\def\NP{Nucl. Phys. B\,}
\def\PL{Phys. Lett. B\,}
\def\PR{Phys. Rev. Lett.}
\def\PRD{Phys. Rev. D\,}
\def\CQG{Class. Quant. Grav.}
\def\IJMP{Int. J. Mod. Phys.}
\def\MPL{Mod. Phys. Lett.}

 
 
\topmargin=.17in                        
\headheight=0in                         
\headsep=0in                    
\textheight=9in                         
\footheight=3ex                         
\footskip=4ex           
\textwidth=6in                          
\hsize=6in                              
\parindent=21pt                         
\parskip=\medskipamount                 
\lineskip=0pt                           
\abovedisplayskip=1em plus.3em minus.5em        
\belowdisplayskip=1em plus.3em minus.5em        
\abovedisplayshortskip=.5em plus.2em minus.4em  
\belowdisplayshortskip=.5em plus.2em minus.4em  
\def\baselinestretch{1.2}       
\thicklines                         
\oddsidemargin=.25in \evensidemargin=.25in      
\marginparwidth=.85in                           
 
 
\def\title#1#2#3#4{
        {\hbox to\hsize{#4 \hfill  #3}}\par
        \begin{center}\vskip.5in minus.1in {\Large\bf #1}\\[.5in minus.2in]{#2}
        \vskip1.4in minus1.2in {\bf ABSTRACT}\\[.1in]\end{center}
        \begin{quotation}\par}
\def\author#1#2{#1\\[.1in]{\it #2}\\[.1in]}

\def\AMIC{Aleksandar Mikovic\'c
\\[.1in]{\it Blackett Laboratory, Imperial College, Prince Consort Road, London
SW7 2BZ, UK}\\[.1in]}

\def\AMICIF{Aleksandar Mikovi\'c\,
\footnote{Work supported by MNTRS and Royal Society}
\\[.1in] {\it Blackett Laboratory, Imperial College, Prince Consort
Road, London SW7 2BZ, UK}\\[.1in]
and \\[.1 in]
{\it Institute of Physics, P.O. Box 57, 11001 Belgrade, Yugoslavia}
\footnote{Permanent address}\\ {\it E-mail:\, mikovic@castor.phy.bg.ac.yu}}

\def\AMSISSA{Aleksandar Mikovi\'c\,
\footnote{E-mail address: mikovic@castor.phy.bg.ac.yu}
\\[.1in] {\it SISSA-International School for Advanced Studies\\
Via Beirut 2-4, Trieste 34100, Italy}\\[.1in]
and \\[.1 in]
{\it Institute of Physics, P.O. Box 57, 11001 Belgrade, Yugoslavia}
\footnote{Permanent address}}

\def\AM{Aleksandar Mikovi\'c 
\footnote{E-mail address: mikovic@castor.phy.bg.ac.yu}
\\[.1in] {\it Institute of Physics, P.O.Box 57, Belgrade 11001, Yugoslavia}
\\[.1in]}

\def\AMsazda{Aleksandar Mikovi\'c 
\footnote{E-mail address: mikovic@castor.phy.bg.ac.yu}
and Branislav Sazdovi\'c \footnote{E-mail: sazdovic@castor.phy.bg.ac.yu}
\footnote{Work supported by MNTRS}
\\[.1in] {\it Institute of Physics, P.O.Box 57, Belgrade 11001, Yugoslavia}
\\[.1in]}

\def\AMVR{Aleksandar Mikovi\'c\,
\footnote{E-mail address: mikovic@castor.phy.bg.ac.yu}
\\[.1in] 
{\it Institute of Physics, P.O. Box 57, 11001 Belgrade, Yugoslavia}
\\[.2in]
Voja Radovanovi\'c \\[.1 in]
{\it Faculty of Physics, P.O. Box 550, 11001 Belgrade, Yugoslavia}}

\def\AMCVR{Aleksandar Mikovi\'c
\footnote{Permanent address: Institute of Physics, P.O. Box 57, 11001 
Belgrade, Yugoslavia}\footnote{E-mail: mikovic@fy.chalmers.se, 
mikovic@castor.phy.bg.ac.yu}
\\
{\it Institute of Theoretical Physics, Chalmers University of Technology,
S-412 96 Goteborg, Sweden}\\[.1in]
and
\\[.1in]
Voja Radovanovi\'c
\footnote{E-mail: rvoja@rudjer.ff.bg.ac.yu} \\
{\it Faculty of Physics, P.O. Box 550, 11001 Belgrade, Yugoslavia}}

\def\AMVVR{Aleksandar Mikovi\'c
\footnote{On leave from Institute of Physics, P.O. Box 57, 11001 
Belgrade, Yugoslavia}
\footnote{Supported by Comissi\'on Interministerial de Ciencia y Tecnologia}
\footnote{E-mail: mikovic@lie1.ific.uv.es}
\\
{\it Departamento de Fisica Te\'orica and IFIC, Centro Mixto Universidad
de Valencia-CSIC, Facultad de Fisica, Burjassot-46100, Valencia, Spain}
\\[.1in]
Voja Radovanovi\'c
\footnote{E-mail: rvoja@rudjer.ff.bg.ac.yu} \\
{\it Faculty of Physics, P.O. Box 368, 11001 Belgrade, Yugoslavia}}

\def\AMV{Aleksandar Mikovi\'c
\footnote{On leave from Institute of Physics, P.O. Box 57, 11001 
Belgrade, Yugoslavia}
\footnote{Supported by Comissi\'on Interministerial de Ciencia y Tecnologia}
\footnote{E-mail: mikovic@lie1.ific.uv.es}
\\
{\it Departamento de Fisica Te\'orica and IFIC, Centro Mixto Universidad
de Valencia-CSIC, Facultad de Fisica, Burjassot-46100, Valencia, Spain}}

\def\endtitle{\par\end{quotation}\vskip3.5in minus2.3in\newpage}
 
 
\def\endabstract{\par\end{quotation}
        \renewcommand{\baselinestretch}{1.2}\small\normalsize}
 
 
\def\xpar{\par}                                         

\def\letterhead{
        \centerline{\large\sf INSTITUTE OF PHYSICS}
        \centerline{\sf P.O.Box 57, 11001 Belgrade, Yugoslavia}
        \rightline{\scriptsize\sf Dr Aleksandar Mikovi\'c}
        \vskip-.07in
        \rightline{\scriptsize\sf Tel: 11 615 575}
        \vskip-.07in
        \rightline{\scriptsize\sf E-mail: MIKOVIC@CASTOR.PHY.BG.AC.YU}}

\def\sig#1{{\leftskip=3.75in\parindent=0in\goodbreak\bigskip{Sincerely yours,}
\nobreak\vskip .7in{#1}\par}}

\def\ssig#1{{\leftskip=3.75in\parindent=0in\goodbreak\bigskip{}
\nobreak\vskip .7in{#1}\par}}

 
\def\ree#1#2#3{
        \hfuzz=35pt\hsize=5.5in\textwidth=5.5in
        \ttraggedright
        \par
        \noindent Referee report on Manuscript \##1\\
        Title: #2\\
        Authors: #3}
 
 
\def\start#1{\pagestyle{myheadings}\begin{document}\thispagestyle{myheadings}
        \setcounter{page}{#1}}
 
 
\catcode`@=11
 
\def\ps@myheadings{\def\@oddhead{\hbox{}\footnotesize\bf\rightmark \hfil
        \thepage}\def\@oddfoot{}\def\@evenhead{\footnotesize\bf
        \thepage\hfil\leftmark\hbox{}}\def\@evenfoot{}
        \def\sectionmark##1{}\def\subsectionmark##1{}
        \topmargin=-.35in\headheight=.17in\headsep=.35in}
\def\ps@acidheadings{\def\@oddhead{\hbox{}\rightmark\hbox{}}
        \def\@oddfoot{\rm\hfil\thepage\hfil}
        \def\@evenhead{\hbox{}\leftmark\hbox{}}\let\@evenfoot\@oddfoot
        \def\sectionmark##1{}\def\subsectionmark##1{}
        \topmargin=-.35in\headheight=.17in\headsep=.35in}
 
\catcode`@=12
 
\def\sect#1{\bigskip\medskip\goodbreak\noindent{\large\bf{#1}}\par\nobreak
        \medskip\markright{#1}}
\def\chsc#1#2{\phantom m\vskip.5in\noindent{\LARGE\bf{#1}}\par\vskip.75in
        \noindent{\large\bf{#2}}\par\medskip\markboth{#1}{#2}}
\def\Chsc#1#2#3#4{\phantom m\vskip.5in\noindent\halign{\LARGE\bf##&
        \LARGE\bf##\hfil\cr{#1}&{#2}\cr\noalign{\vskip8pt}&{#3}\cr}\par\vskip
        .75in\noindent{\large\bf{#4}}\par\medskip\markboth{{#1}{#2}{#3}}{#4}}
\def\chap#1{\phantom m\vskip.5in\noindent{\LARGE\bf{#1}}\par\vskip.75in
        \markboth{#1}{#1}}
\def\refs{\bigskip\medskip\goodbreak\noindent{\large\bf{REFERENCES}}\par
        \nobreak\bigskip\markboth{REFERENCES}{REFERENCES}
        \frenchspacing \parskip=0pt \renewcommand{\baselinestretch}{1}\small}
\def\unrefs{\normalsize \nonfrenchspacing \parskip=medskipamount}
\def\Item{\par\hang\textindent}
\def\Itemitem{\par\indent \hangindent2\parindent \textindent}
\def\makelabel#1{\hfil #1}
\def\topic{\par\noindent \hangafter1 \hangindent20pt}
\def\Topic{\par\noindent \hangafter1 \hangindent60pt}

\title{Quantum corrections for the \RN charged black hole}
{\MBVR}{}{July 1999}

\noindent
We calculate the quantum corrections of geometric and thermodynamic
quantities for the \RN charged black hole, within the context of 2D
spherically symmetric dilaton gravity model. Special attention is payed
to the quantum corrections of the extreme \RN solution. We find a state
of the extreme black hole with regular behaviour at the horizon.

\endtitle 

\section{Introduction}

Many aspects of the radiation of black holes were
investigated  since its discovery in 1975 \cite{h}.
The work which has been done in the last years 
is mainly done in the framework  of
various dilaton two dimensional (2D) models, 
due to renewed interest in 2D gravity
originated in the string theory. In this context, 
two main types of models have been analyzed.
One of them are string-inspired models with black-hole solutions
(CGHS, RST, BPP) which are exactly solvable.
 The motivation to analyze the other class of 2D models
like  null-dust model, spherically symmetric gravity (SSG), etc., is that
 they are obtained from 4D gravity action by some
variant of dimensional reduction and it is expected that they can
reproduce some  aspects of 4D gravitational phenomena.

In this paper we analyze the quantum corrections to the \RN black hole in
the framework of SSG. 
Extreme black holes have an important place in some recent
investigations. First, as extreme black hole is the
solution with  critical behaviour \cite{Cai}, i.e. a sort of phase
transition, it obeys some scaling laws. This enables one to discuss such
questions as  problem of information loss, avoiding the limits which
the quanization of gravity (issues of the existence of Planck
length) puts. Second, extreme dilaton black holes are representations
of massive single string states in superstring theories while extreme
black p-branes occur as solutions in M-theory; therefore it is of
importance to analyze the properties of these and similar solutions.

One of the extreme black hole solutions is the Reisner-Nordstr\"om
charged black hole. There are two main streams of thought
about the thermodynamic properties of this solution. Hawking
\cite{HHR} argues that, as this solution has a double zero on the
horizon and therefore does not have a conical singularity (in its
Euclidean exstension), one can prescribe it arbitrary temperature.
 This means that extreme black hole can be in equilibrium with
its radiation at arbitrary temperature. Furthermore, this property
implies that the entropy of extreme black hole is zero, i.e. it is
not equal to the fourth of its area as it would be according to the
Bekenstein-Hawking formula. This point of view was further reinforced by the
discussion of Teitelboim \cite{T}, which partially also implies that the extreme
solution is stable, as it has  different topology from nonextreme
solutions. On
the other hand, the calculations of Trivedi \cite{trivedi} and
Anderson, Hiscock and Loranz
\cite{lha,ahl} show that the extreme black hole at  nonzero
temperature cannot exist (is not stable), because the energy-momentum tensor
(EMT)  of the radiation diverges on the horizon
 in the semiclassical approximation.
There is a further discussion about stability at 
zero temperature: Trivedi obtained
that, in his particular 2D model, EMT diverges for zero temperature 
on the horizon also,
but "mildly", in
such a way that the curvature and the tidal forces remain finite. The
calculations of \cite{lha,ahl} suggest that in 4D and $T=0$, EMT is finite
on the horizon. So, as it can be seen,
the questions of temperature and entropy of extreme \RN black hole
are still unsettled.

In our previous paper \cite{brm},
 we calculated the backreaction of the radiation on the
Schwarzschild solution in the framework of the SSG dilaton gravity model.
The effective action which was used to describe the one-loop quantum effects
was found in \cite{mr4,bh,no,klv}. In the SSG 
model the coupling of the scalar field to
gravity is supposed to be more realistic (from the point of view of four
dimensions), because
it is obtained from the 4D minimal coupling
of scalar field to gravity. This is new in comparison
to the previous calculations \cite{trivedi}, and it is
therefore interesting to see
whether it will produce also some differencies in the  conclusions
about the behaviour of  the \RN solution. Also,
our approach for finding the value of EMT and identification
of \HH vacuum proves to be relatively simple in the \Sch case,
 which is a further
motivation to see what can this model  tell about the behaviour of the
extreme solution.
Indeed, an unexpected result comes out:  within this model there is
a Hartle-Hawking vacuum with  fixed (possibly nonzero)
 temperature, defined in
 the sense that the
semiclassical value of the EMT tensor is regular
everywhere (except at the physical singularity $r=0$).
Unfortunately, the temperature of this state cannot be
determined, as the scalar field is not free in the asymptotic region
and therefore the Stefan-Boltzmann law cannot be applied. 
The question of stability of this state is also an interesting
point to consider.

The plan of this paper is the following: In the second section we define
the model and obtain the quantum corrections for the nonextreme \RN
solution. In the third section we discuss properties of
the extreme \RN solution. 
The analysis of
 the possibility to define the  \HH
vacuum and the Kruscal coordinates is also given.
A generalization of our model to the model with $\xi Rf^2$ coupling and
some results are given in the Appendix.

\section{The Model and its Solution}

The starting  point of our consideration  is the 4D action
\be \label{eq:S4}
\G_0={1\over 16\p G}\int d^4x\sqrt{-g^{(4)}}(R^{(4)}-F^2)-{1\over
8\pi }\sum _i\int d^4x\sqrt{-g^{(4)}}((\nabla f_i)^2+\xi Rf_i^2)\ ,\ee
which describes
the Einstein gravity coupled with electromagnetic field $A_\mu$ and scalar
fields $f_i$ ($i=1,\dots N$). The $\xi Rf_i^2$ interaction
term is added  to the action. 
Specially, if $\xi ={1\over 6}$, the second term in
(\ref{eq:S4}) is conformally invariant. Using the equation of motion
for electromagnetic field and the solution
which corresponds to the Coulomb potential (as
it is done in \cite{fis}) and performing the
spherically symmetric reduction, from action (\ref{eq:S4})
we get 2D action
\bea \G_0&=&{1\over 4G}\int d^2x\sqrt{-g}
\Bigl( e^{-2\Phi}(R+2(\nabla\Phi)^2+2e^{2\Phi}(1-Q^2e^{2\Phi}))\ncr
&-&
2Ge^{-2\Phi}\sum _i\bigl((\nabla f_i)^2+\xi f_i^2(R-6(\nabla\Phi)^2+
4\Box\Phi+2e^{2\Phi})\bigr) \Bigr)  \ ,\label{eq:S2} \eea
where $\Phi$ is the dilaton field and $Q$ is the electric charge.
When we add the one-loop quantum correction for the matter fields $f_i$
 to the  action (\ref{eq:S2}), we get
the nonlocal effective action. It is given by (see Appendix):
\bea \label{eq:S2eff} \G_1&=&{1\over 4G}\int d^2x\sqrt{-g}
\Bigl( e^{-2\Phi}(R+2(\nabla\Phi)^2+2e^{2\Phi}(1-Q^2e^{2\Phi}))\ncr
&-&
2Ge^{-2\Phi}\sum_i\bigl( (\nabla f_i)^2+\xi f_i^2(R-6(\nabla\Phi)^2+
4\Box\Phi+2e^{2\Phi})\bigr)\Bigr) \ncr
&-&
{N\over 8\pi}\int d^2x\sqrt{-g}\Bigl( ({1\over 12}-\xi )R{1\over \Box}R+
(1-4\xi )R\Phi \ncr
&+&
(-1+6\xi )R{1\over\Box }(\nabla\Phi )^2-2\xi
R{1\over\Box}e^{2\Phi}\Bigr)\ . \eea

The effective action (\ref{eq:S2eff})
can be rewritten in the local form by introducing two
auxillary fields, $\psi $ and $\chi \ (\xi
\neq {1\over 12},\ {1\over 6})$
\bea \label{eq:S2aux}
\G_1&=&{1\over 4G}\Big[ \int d^2x\sqrt{-g}\Bigl(
 e^{-2\Phi}(R+2(\nabla\Phi)^2+2e^{2\Phi}(1-Q^2e^{2\Phi}))
\ncr
&-& 2Ge^{-2\Phi}\sum_i\bigl(\nabla f_i)^2+\xi f_i^2(
R-6(\nabla \Phi)^2+4\Box \Phi +2e^{2\Phi})\bigr)\Bigr) \ncr
&-&
\kappa (1-12\xi )\int d^2x\sqrt{-g}\Bigl( (\nabla\psi )^2+2R\psi +
b(\nabla\psi )(\nabla\chi )\ncr
&+&b\psi(\nabla \Phi)^2+bR\chi +c\psi e^{2\Phi}+dR\Phi
\Bigr)\Big]\ .  \eea
In (\ref{eq:S2aux}) the following notation is introduced:
$\kappa ={GN\over 24\pi}$,
$b=-12{1-6\xi\over 1-12\xi }$, $c={-24\xi\over 1-12\xi }$,
$d=12{1-4\xi\over 1-12\xi}$.
The auxilliary fields   $\psi$, $\chi$ obey the equations of motion:
\be \Box\psi =R \label{eq:psi} \ee
\be \Box\chi =(\nabla\Phi )^2+{c\over b}e^{2\Phi} \label{eq:chi} \ee
Substitution of (\ref{eq:psi}) and (\ref{eq:chi}) into (\ref{eq:S2aux})
gives the nonlocal action (\ref{eq:S2eff}).
In the case $\xi ={1\over 12}$, the local form of the one-loop
correction is given by
\be
\G^{(1)}={N\over 48\p}\int d^2x\sqrt{-g}\Big(R\chi +(3(\nabla \Phi )^2+e^{2\Phi}
)\psi +\nabla \psi\nabla \chi-4R\Phi \Big) \ ,
\ee
where the auxilliary fields obey $\Box \psi=R, $ and $\Box \chi
 =3(\nabla \Phi )^2+e^{2\Phi}$. In the other special case $ \xi
={1\over 6}$  we get
\be
\G^{(1)}={N\over 48\p}\int d^2x\sqrt {-g}\Big( R\psi +2R\chi +2\psi
e^{2\Phi}+{1\over 2}(\nabla \psi)^2+2\nabla \psi \nabla \chi -2R\Phi
\Big) \ ,
\ee
where
$\Box \psi =R\ ,\Box \chi =e^{2\Phi}.$

The action (\ref{eq:S2eff}) in the classical limit $\kappa =0$
posesses
the Reisner-Nordstr\"om black hole as a vacuum solution.
Since the scalar fields $f_i$
enter the action (\ref{eq:S2eff}) quadratically, we can get the
desired quantum correction to the vacuum solution
by introducing $f_i=0$ directly to the action and then obtaining the equations
of motion. Furthermore, we will take also $\xi = 0$ for simplicity, 
and discuss the case
$\xi\neq 0$ briefly in the Appendix. The action which we will vary is
\bea \label{eq:S} \G _1&=&{1\over 4G}\int d^2x\sqrt{-g}\Bigl( r^2R+2(\nabla
r)^2+2U(r)\Bigr) \ncr
&-& {\kappa\over 4G}
 \int d^2x\sqrt{-g}\Bigl( (\nabla\psi )^2+2R\psi -12(\nabla\psi )
(\nabla\chi ) \ncr
&-&12\psi{(\nabla r)^2\over r^2}-12R\chi -12R\log{r}
\Bigr)\Bigr], \eea
where, instead of the dilaton $\Phi$ we introduced the field
$r=e^{-\Phi}$ which has the meaning of radius.
$U(r)=1-{Q^2\over r^2}$ is the dilaton potential. From the action
 (\ref{eq:S}) we get the following equations of motion:

\be {\Box\psi =R}\ee

\be {\Box\chi ={(\nabla r)^2\over r^2} }\ee

\be{ 2\Box r-rR-U^\prime = -6 \kappa \Big( 2\psi {\Box r\over r^2}
+2{(\nabla\psi )(\nabla r)\over r^2}-2\psi {(\nabla r)^2\over r^3}+
{R\over r}\Big) } \label{eqr}\ee

\bea &&g_{\mu\nu}\big( \Box r^2-(\nabla r)^2-U\big) -2r\nabla _\mu\nabla _\nu r=
 2G  T_{\mu\nu}=\ncr
&=&\kappa \Bigl( g_{\mu\nu}(2R+6\psi {(\nabla r)^2\over r^2}-{1\over 2}
(\nabla\psi )^2+
6(\nabla\psi )(\nabla\chi )-12{\Box r\over r}) \ncr
&+&\nabla _\mu\psi\nabla _\nu\psi -12\nabla _\mu\psi\nabla _\nu\chi
-2\nabla _\mu\nabla _\nu\psi +12\nabla _\mu\nabla _\nu\chi  \ncr
&+&12{\nabla _\mu\nabla _\nu r\over r}
-12(1+\psi ){\nabla _\mu r\nabla _\nu r\over r^2} \Bigr)\ . \label{eqg} \eea

The static classical solution
for the dilaton and metric is given by
$$r=x^1\ ,$$
$$g_{\mu\nu}=\pmatrix{ -f_0 & 0 \cr 0 & {1\over f_0} \cr}\ ,$$

$$f_0={(r-{r_+})(r-{r_-})\over r^2} = 1-{2MG\over r}+{Q^2\over r^2}\ . $$
This is the well known \RN  solution
 which describes charged black
hole of mass $M$ and charge $Q$.
The horizons of
the black hole are $r_\pm =MG\pm \sqrt{(MG)^2-Q^2} $.
The static solution  for the
auxilliary fields in the
zero'th  order is
\bea \label{ps} 
\psi _0&=&Cr+2\log {{r\over l}}+{-{r_+}+{r_-}+C{r_+}^2\over {r_+}-{r_-}}
\log{{r-{r_+}\over l}}\ncr
&+&{-{r_+}+{r_-}-C{r_-}^2\over {r_+}-{r_-}}\log{{r-{r_-}\over l}}\ ,\eea

\bea \chi _0&=&Dr-{1\over 3}\log{{ r\over l}}
+{-3{r_+}+6D{r_+}^2+{r_-}\over 6({r_+}-{r_-})}\log{{r-{r_+}\over l}}\ncr
&+& {-{r_+}+3{r_-}-6D{r_-}^2\over 6({r_+}-{r_-})}\log{{r-{r_-}\over
l}}\ . \label{ch} \eea
In order to have dimensionless expressions under the logarithm, we introduced
one integration constant $l$ of the dimension of length in both $\psi _0$ 
and $\chi _0$. The constants $C$ and $D$
are introduced similarly as in the Schwarzchild case \cite{brm},
and they are of great importance because
they describe the quantum state of matter. We are
interested to fix them in accordance with the
Hartle-Hawking boundary conditions.

 We want to calculate the one-loop quantum correction to the given classical
solution.
The static ansatz is the following:
\be r=x^1 \ ,\ee
\be g_{\mu\nu}=\pmatrix{ -fe^{2\phi} & 0 \cr 0 & {1\over f} \cr}\ee
where
$$f={(r-{r_+})(r-{r_-})\over r^2}+\kappa{m(r)\over r}\ ,$$
and
$$\phi (r) = \kappa\varphi (r)\ .$$

We can easily solve the equations of motion (\ref{eqr}-\ref{eqg})
in the first order in $\kappa$.
The unknown functions $m(r)$ and $\varphi (r)$ in this order satisfy

\be m^\prime =-{2G\over \k}{1\over f}T_{00}\label{eq:eqm} \ee
\be\varphi ^\prime ={2G\over \k}{1\over 2r}(T_{11}+{T_{00}\over f^2})\ .
\label{eq:eqfi} \ee
The  equations (\ref{eq:eqm}) and (\ref{eq:eqfi})
can easily be integrated, but the solution  
for $m(r)$ and $\varphi (r)$ is so long that
it makes no sense to write it in full length.
We will first  determine
 the values of the
 constants $C$ and $D$ using the strategy of \cite{brm}.
Both $m$ and $\varphi$ are singular at the
horizons $r_+$ and ${r_-}$: $m$ has logarithmic singularities, while
 $\varphi$ has
 logarithmic and ${1\over r}$
singularities. Condition that the coefficients of all  singular
terms at the outer horizon ${r_+}$ vanish gives a relation
between  $C$ and $D$:
\be 6C{r_+}^3+C(C-12D){r_+}^4+2{r_+}{r_-}-{r_-}^2-{r_+}^2(1+2C{r_-})=0\ .
\label{eq:U1} \ee
The other relation comes from the condition of regularity
of the first correction of 
curvature at $r_+$,
$$R_1=-\left({m\over r}\right)^{\prime\prime}
-{f_0}^\prime\varphi ^\prime -2f_0
\varphi ^{\prime\prime}\ ,$$
and it is given by
\be -18C{r_+}^3-C(C-12D){r_+}^4-26{r_+}{r_-}+13{r_-}^2+{r_+}^2(1+2C{r_-})=0
\  .
\label{eq:U2} \ee
The solution of (\ref{eq:U1}) and (\ref{eq:U2}) is
\be \label{eq:CD}
C={{r_+}-{r_-}\over {r_+}^2}\ ,\ \ D={3{r_+}-{r_-}\over 6{r_+}^2}\ .\ee
Note that for $Q=0$ ($r_- =0$)
the values of constants $C$ and $D$ are the same as in \cite{brm}; for
$r_-=r_+$ we have $C=0$.

Introducing the values (\ref{eq:CD}) in the solution for $m$ and $\varphi$,
 we get the quantum corrections of the metric tensor $g_{\mu\nu}$:
\bea \label{eq:m} m(r)=&-&{1\over 2{r_+}^4{r_-}^2r^3}\Bigl[
 -r^4(5{r_+}^2-6{r_+}{r_-}+{r_-}^2){r_-}^2\ncr
&-&r^2({r_+}^3-4{r_+}^2{r_-}+{r_+}{r_-}^2-6{r_-}^3){r_+}^2{r_-}\ncr
&-&2r(3{r_+}^2+{r_+}{r_-}+2{r_-}^2){r_+}^3{r_-}^2
+4{r_+}^5{r_-}^3\ncr
&+&\log {r\over r-r_-}\,\Bigl( 12r_+^4r_-^2r^2-
6(r_++r_-)r_+^4r_-^2r+4r_+^5r_-^3+r_+^2(r_+^3-5r_+^2r_-)r^3\Bigr)\ncr
&+&\log {r\over l}\,\Bigl(
12r_+^4r_-^2r^2-6(r_++r_-)r_+^4r_-^2r+4r_+^5r_-^3+r_+^2(r_-^3-5r_+r_-^2)r^3
\Bigr) \ncr
&-&\log {r-r_-\over l}\,\Bigl( (r_-^5-5r_+r_-^4)r^3+12r_+^2r_-^4r^2-6(r_++r_-)r_+^2r_-^4r+
4r_+^5r_-^3\Bigr) \Bigr]\ncr
&+& m_0
\eea
and

\be \varphi (r)=F(r)-F(L)\ ,\ee
where
\bea F (r)=&-&{1\over 2{r_-}^2{r_+}^4r^2(r-{r_-})}\Bigl[
2{r_+}^4{r_-}^3+2r(4{r_+}-3{r_-}){r_+}^2{r_-}^3 \ncr
&-&r^2(3{r_+}^4+8{r_+}^3{r_-}-4{r_+}^2{r_-}^2-4{r_+}{r_-}^3
+{r_-}^4){r_-}   \Bigr]\ncr
 &+&\log{r\over r-r_-}\, ({3\over r^2}+{1\over r_+r_-}-{3\over 2r_-^2})
 +\log{r\over l}\, ({3\over r^2}-{3\over 2r_+^2})\ncr
 &+&\log {r-r_-\over l}\, ({3r_-\over r_+^3}-{3r_-^2\over r_+^2r^2}-{r_-^2\over
2r_+^4}-{1\over r_+^2})\ .
 \eea
 This quantum correction, as we shall see, defines the \HH vacuum state.
In the solution for $m$ we entered  arbitrary integration constant
$m_0$. This constant
 is in principle related to the redefinition of the mass of
black hole and we want to keep it to the end of calculations in order
to check whether it influences the results. $L$ is introduced to
describe the size of our system; as in \cite{y}, we are considering
black hole in thermodynamic equilibrium with radiation in the box of
size $L$ with ideally reflecting walls.

  Let us rewrite the zero'th order solution for the auxilliary fields
after the introduction of values (\ref{eq:CD})  for $C$, $D$:
\be \psi _0={r_+-r_-\over {r_+}^2}r+2\log{{r\over
l}}-{{r_+}^2+{r_-}^2\over {r_+}^2}\log{{r-r_-\over l}}\ ,\ee
\be \label{rechi} \chi _0={3r_+-r_-\over 6{r_+}^2}r-{1\over 3}\log{{r\over
l}}-{({r_+}-{r_-})^2\over 6{r_+}^2}\log{{r-r_-\over l}}\ .\ee
From the solutions (\ref{eq:m}-\ref{rechi})
we see that the  singularities on the inner horizon persisted, while 
singularities on the outer horizon vanished for the auxilliary fields, too.

If we define the null-coordinates for the \RN black hole as usual 
\cite{chan, he}:
$u=t-r_*$, $ v=t+r_*$, where the "tortoise coordinate" $r_*$ is given by
\be r_*=r+{r_+^2\over r_+-r_-}\log (r-r_+)-{r_-^2\over r_+-r_-}\log
(r-r_-)\ ,\ee
the metric takes the form
\be ds^2=-\big( 1-{2M\over r}-{Q^2\over r^2}\big) dudv. \ee

The ingoing and outgoing fluxes are given by
\bea T_{uu}&=&T_{vv}={1\over 48\p}{(r-{r_+})^2\over 4r^6{r_+}^4}\Bigl[
8{r_+}^4{r_-}^2
-4r(3{r_+}^2+2{r_+}{r_-}-3{r_-}^2){r_+}^2{r_-}\ncr
&+& 3r^2({r_+}-{r_-})({r_+}+7{r_-}){r_+}^2
-2r^3(5{r_+}^2-6{r_+}{r_-}+{r_-}^2){r_+}\ncr
&-&r^4(5{r_+}^2-6{r_+}{r_-}+{r_-}^2)\ncr
&+&12{r_+}^2(r-{r_-})^2\Big( ({r_+}^2+{r_-}^2)
\log {r-{r_-}\over l}-2r_+^2\log{r\over l} \Big) \Bigr] \ ,
\label{eq:EMT}\eea
while
\be \label{eq|EMT1} T_{uv}={1\over 48\p r^6}(r-{r_+})(r-{r_-})
(-3r_+r_-+2rr_++2rr_-)
\ .\ee

In order to investigate the regularity of EMT on the outer horizon,
we have to use the free-falling observer frame \cite{ahl}.
The  coordinates which describe this frame 
are the Kruscal  $\{U,V\}$-coordinates.
 They are \cite{he}
\be \label{eq:Kr} U=-e^{-\a u},\ \ \ V=e^{\a v}\ ,\ee
where
\be \alpha ={r_+-r_-\over 2r_+^2}\ .\ee
The components of EMT in the Kruscal coordinates take the form
\be T_{UU}={1\over \a^2U^2}\, T_{uu}={V^2\over \a^2}e^{-4\a
r}( r-r_-)^{{2r_-^2\over r^2_+}}{T_{uu}\over (r-r_+)^2}\ee
\be T_{VV}={1\over \a^2V^2}\, T_{vv}\ee
\be T_{UV}=-{1\over \a ^2UV}\, T_{uv}={1\over \a^2}e^{-2\a
r}(r-r_-)^{{r_-^2\over r_+^2}}{T_{uv}\over r-r_+}\ .\ee
So, from (\ref{eq:EMT}) and (\ref{eq|EMT1}) we
easily see that on the horizon
$r=r_+$, i. e. $V=0, U=$const,
the regularity conditions
\be T_{UU}<\infty,\ T_{VV}<\infty, \ T_{UV}<\infty \ee
or, equivalently,
\be { T_{uu}\over f^2}<\infty,\ T_{vv}<\infty, 
\ {T_{uv}\over f}<\infty \ ,\label{frame} \ee
are fulfilled for the values of constants $C$ and $D$ given by (\ref{eq:CD}).
These values of
constants define that 
our sistem is in the thermal Hartle-Hawking state. 

It is straightforward to confirm this statement also directly,
along the lines of Balbinot, Fabbri,
transforming the value of EMT from the Boulware state
to the \HH   conformal  $\sket{U,V}$ state. Namely, in accordance with
\cite{bf}, the values of the EMT tensor in the null-coordinates in
the Boulware state for the action (\ref{eq:S}) are given by
\bea \bra{B}\,\hat T_{uv}\,\ket{B}&=&-{1\over 12\p}
(\partial _+\partial _-\rho +3\partial
_+\Phi\partial _-\Phi -3\partial _+\partial _-\Phi )\ncr
&=&{1\over 48\p}({1\over 2}f\, f^{\prime\prime}+{3\over r}f\,
f^\prime)\ ,\eea
and
\bea \bra{B}\,\hat T_{uu}\,\ket{B}&=&\bra{B}\,{\hat T_{uu}}^{{\rm PL}}\,\ket{B}
+{1\over 2\p}
\Big( \rho(\partial _- \Phi )^2+{1\over 2}{\partial _-\over\partial _+}
(\partial _+\Phi \partial _-\Phi )\Big) \ncr
&-&{1\over 4\p}( -2(\partial _-\rho)(\partial _-\Phi )+
(\partial _-\Phi )^2)\ncr
&=&\bra{B}\,{\hat T_{uu}}^{{\rm PL}}\,\ket{B}
+{1\over 16\p}{f^2\over r^2}\log f ,\eea
where $\rho ={1\over 2}\log f$ is the conformal factor and
$f=f_0={(r-r_-)(r-r_+)\over r^2}$.
$\bra{B}\,{\hat T_{uu}}^{{\rm PL}}\,\ket{B}$  
is the value of EMT for the case where the
complete effective action consists of the Polyakov-Liouville term only.
The conformal transformation to the
\HH state $\sket{U,V}$ gives for the value of EMT 
\be \bra{H}\,\hat T_{uv}\,\ket{H}=\bra{B}\,\hat T_{uv}\,\ket{B}\ee
\bea \bra{H}\,\hat T_{uu}\,\ket{H}&=&\bra{B}\,\hat T_{uu}\,\ket{B}
+{1\over 24\p}({F^{\prime\prime}\over F}
-{1\over 2}{F^{\prime 2}\over F^2})\ncr
&+&{1\over 4\p}\Big( (\partial _-\Phi )^2\log FG+{F^\prime\over F}
\int (\partial _+\Phi )(\partial _-\Phi ) dv \Big) \ . \label{eq:EMTbf} \eea
Here, $F(u)={du \over dU}$, $G(v)={dv\over dV}$. It if easy to check
that, when calculated, (\ref{eq:EMTbf}) really coincides with the
value (\ref{eq:EMT}) obtained previously.

Knowledge of the functions $m$ and $\varphi$ allows us to calculate the
first quantum corrections of the positions of the horizons, temperature and entropy
of the charged black hole.
The values (\ref{eq:CD})
assure regularity of these corrections, as we shall shortly see.
 If we define  the corrected positions
of inner and outer horizons as
$$r_+^q={r_+}+\kappa r_1\ ,\ \ r_-^q={r_-}+\kappa r_2\ ,$$
we get
$ r_1 =-{{r_+}\over {r_+}-{r_-}}m({r_+})$,
 $ r_2 =-{{r_-}\over {r_-}-{r_+}}m({r_-})$.
Using (\ref{eq:m}), we obtain
\bea r_1&=&-{1\over 2{r_+}^3{r_-}^2}\Big[ {r_+}{r_-}({r_+}^2+8{r_+}{r_-}
+{r_-}^2) -(r_+^4+2r_+^2r_-^2-4r_+^3r_-)\log {r_+\over r_+-r_-} \ncr
&-&{r_+}^2{r_-}^2 \log{r_+\over l}+{r_-}^4\log{{r_+}-{r_-}\over l}\Big]
-{r_+\over r_+-r_-}m_0\ \eea
for the correction of the outer horizon.

The Hawking temperature is defined by
$$ T={1\over 4\p}e^\phi f^\prime\vert _{r_+^q}=T_0+\k T_1$$
where, as it is known,  $T_0={1\over 4\p}{r_+-r_-\over r_+^2}$. The first
 correction  of temperature $ T_1$ in the case of nonextreme
 \RN solution is given by

 \bea  4\p T_1&=&{{r_+}-{r_-}\over {r_+}^2}\phi ({r_+})+{m^\prime
({r_+})\over r_+}+{m({r_+})(r_+-3r_-)\over {r_+}^2(r_+-r_-)}= \ncr
&=&
{1\over 2{r_+}^6{r_-}}\Bigl( 4{r_+}^4+5{r_+}^3{r_-}-33{r_+}^2{r_-}^2-{r_+
}{r_-}^3+{r_-}^4\Bigl) \ncr
&+& ({4\over r_+^3}+{2\over r_+r_-^2}-{6\over r_+^2r_-})\log
{r_+-r_-\over r_+}\ncr
&+&({4r_-\over r_+^4}-6{r_-^2\over
r_+^5}+2{r_+^3\over r_-^6})\log{r_+-r_-\over l}\ncr
&-&{r_+-r_-\over r_+^2}F(L)+
{m_0(r_+-3r_-)\over r_+^2(r_+-r_-)}\ .
\eea

The standard definition of entropy, based, e.g., on the conical
singularity method \cite{brm}:
\be S={\pi\over G}(r^2-\kappa (2\psi -12\chi -12\log r))\vert _{ r_+^q}=
S_0+\kappa S_1 \ ,\label{eq:ent} \ee
 gives
for its correction linear in $\kappa$
\bea S_1&=&{\pi\over G}
\Bigl[  {2m_0\over r_+(r_+-r_-)}-4-{r_+\over r_-}-{r_-\over r_+}
+5\log {r_+\over l} \ncr
&+&(2+{r_+^2\over r_-^2}-{4r_+\over r_-})\log {r_+\over r_+-r_-}+
({4r_-\over r_+}-{r_-^2\over r_+^2})\log{r_+-r_-\over l}\Bigr] ,\eea
up to constants. This represents the quantum correction to the standard
Bekenstein-Hawking entropy.

\section{Extreme black hole}

Extreme \RN black hole  is defined by the
 coincidence of the inner and outer horizons, ${r_-}={r_+}$.
In this case the function $f(r)$ which defines the metric has a double zero
at $r=r_+=MG$.
Notice that, technically,
this implies difference in
the calculations which were performed,  e.g.
the types of singularities which we obtained are different.

Let us assume
first that the extreme black hole can be approached in
the limit of nonextreme
ones, e.g. in a way similar to that of Zaslavskii \cite{z1,z2}.
We shall assume that this means  taking the
limit ${r_-}\to {r_+}$ in the results that are
 already obtained for the nonextreme
 solutions, i.e. at the end of calculations.
 Let us notice that this fixes the value of the constant
C in the auxilliary field $\psi$ to zero:
$C={{r_+}-{r_-}\over {r_+}^2}=0$, while for $D$ we get
$D={1\over 3r_+}$. The values of
$C$ and $D$ give the change of the metric at infinity, telling us what
is the temperature (energy) of the Hawking radiation. $C=0$ corresponds
to the case that the metric remains Minkowskian in the asymptotic region, i.e.
that the temperature of  radiation is zero.
So, extreme limit taken in this sense really gives zero temperature.

The functions $m$, $\varphi$ and $ F$ which we have calculated
 in this case reduce to:

\be \label{eq:mex} m(r)={2(r-{r_+})
\over r^3{r_+}}\Bigl[ {r_+}({r_+}-2r)+
2(r-{r_+})^2\log{{r\over r-{r_+}}}\Bigr] +m_0 \ee
\be F(r)={1\over r^2{r_+}^2}\Bigl[ {r_+}(2r+{r_+})-
2(r^2-3{r_+}^2)\log {{r\over r-{r_+}}}\Bigr] \ .\ee
The first quantum correction of the curvature is
\be \label{eq:Rex} R_1(r)=-{4\over r^6{r_+}}\Bigl[ {r_+}(-2r^2+19r{r_+}
-19{r_+}^2)+2(r^3-9r{r_+}^2+8{r_+}^3)\log{{r\over r-{r_+}}}\Bigr] \ .\ee
From (\ref{eq:mex}-\ref{eq:Rex})
we see that $m$ and $R_1$ are not singular at the horizon, while
 $\varphi$ is.

 The corrections of the position of the horizon which we get in
this case are
\be r_1=-{5\over {r_+}}\ , r_2={5\over {r_+}}\ .\ee
if we assume that the constant $m_0=0$; otherwise the result is divergent.
From the last expression we can see that classical extreme black hole
would turn to the nonextreme
when quantum corections are taken, i. e. $r_+^q\neq r_-^q$.

For the correction of the temperature we get
\be 4\pi T_1=-{12\over {r_+}^3}\ ,\ee
for $m_0=0$.
This would mean that the quantum corrected temperature is negative,
because, as
we have already seen, the classical temperature is zero,
$T_0=0$. This result is not physically acceptable.
The one-loop correction of the entropy
$$ S_{1}={\pi\over G}\big[ -6+4\log {r_+( r_+-r_-)\over l^2}
\big]\vert _{r_-=r_+}  $$
is divergent and therefore  also difficult to interpret.

To conclude the discussion of this case, in order to compare our
results with \cite{trivedi,lha}, let us analyze the behavior of the EMT.
The nonsingularity of EMT  in
the free-falling observer frame \cite{lha} gives that
$T_{vv}$, $T_{uv}/f$ and $T_{uu}/f^2$
are nonsingular on the horizon. For $T_{uu}$ we get

\be T_{uu}=T_{vv}=-{(r-{r_+})^3
\over 24\p r^6}({r_+}+3(r-{r_+})\log{ {r\over r-{r_+}}}) \label{eq:t} \ee
while
\be T_{uv}={r_+(r-r_+)^2(4r-3r_+)\over 48\p r^6} \ .\label{t1} \ee

Since the metric factor is  $f={(r-{r_+})^2\over r^2}$,
we see that the EMT is  divergent at the horizon unlike the nonextreme case,
 as Trivedi
also notices \cite{trivedi}. This result implies that the state
defined by the choice $C=0, D={1\over
3r_+} $ is not thermal.
 Let us also note that the values of EMT (\ref{eq:t}), (\ref{t1})
coincide with the values of EMT for the Boulware state, $\bra{B}\,
\hat T_{uu}\,\ket{B}$
and $\bra{B}\,\hat T_{uv}\,\ket{B}$, calculated in accordance to \cite{bf}.

Let us finally 
note that, as our results also show, the "extreme limit" in the
sense which we have just discussed  could not exist in reality, i.e.
in physical experiment.
This is because the temperature of
nonextreme black holes decreases towards zero in the 
limit $MG\to Q$, so in thermodynamic sense
 approaching the extreme black hole
would be  the same as approaching the absolute zero.

Now we turn to the other possibility, namely  finding the quantum
corrections
of the extreme black hole solution from the beginning. This procedure is
justified by the fact that
the topology of extreme solution 
is different from the topology of nonextreme one \cite{T}. 
The ansatz for the correction is
$$r=x^1\ ,$$
$$g_{\mu\nu}=\pmatrix{ -fe^{-2\Phi} & 0 \cr 0 & {1\over f} \cr}$$
where now the function $f$ is given by
$$f={(r-{r_+})^2\over r^2}+\kappa{m(r)\over r}\ .$$
The zero'th order solutions for
 the auxilliary fields are
\be \psi _0=Cr-{C{r_+}^2\over r-{r_+}}+2\log {r\over l}+2C{r_+}
\log{{r-{r_+}\over l}}\ ,\label{psi}\ee

\be \chi _0=Dr-{1\over 3}\log {r\over l}+{{r_+}(1-3D{r_+})
\over 3(r-{r_+})}-{2(1-3D{r_+})\over 3}\log{r-{r_+}\over l}\label{chi}
\ .\ee
Note that (\ref{psi}) and (\ref{chi}) are not the same as (\ref{ps}) and 
(\ref{ch}) in the limit $r_+\to r_-$. 
Proceeding on the similar lines as in the previous section,
we can solve the equations for $m$ and $\varphi$ and impose the
conditions of regularity.
It is interesting that now also  a choice of
constants $C$ and $D$ exists which enables the regularity
 $m$, $\varphi$, $R_1$ on the horizon, and EMT in the free falling
observer frame.
It is given by
\be \label{eq:CDex} C={1\over {r_+}}\  , D={5\over 12{r_+}}\ .\ee
For $m$ and $\varphi$ in this case we get
\be m(r)={2\over {r_+}^2r^3}\Big[ r^4+4{r_+}^3r-2{r_+}^4+2r_+(r-{r_+})^3\log
{r\over l}\Big] +m_0 \ee
\be\varphi (r) =F(r)-F(L)\ ,\ee
where
\be F(r)={2\over {r_+}^2r^2}\Big[ 2{r_+}( 2r+{r_+})-(r^2-3{r_+}^2)\log
{r\over l}\Big] \ .\ee
The auxilliary fields become
\be \psi (r)={r\over {r_+}} -{{r_+}\over r-{r_+}}+2\log {r\over l}\ , \ee
and
\be \chi (r)={1\over 12}\Big[ {5r\over {r_+}}-{{r_+}\over r-{r_+}}
-4\log {r\over l}+2\log {r-{r_+}\over l} \Big]\ .\ee
We see that for
the values (\ref{eq:CDex}) the auxilliary
fields diverge at $r=r_+$, unlike $m$ and $\varphi$.

It is easy to see that the equation defining the
position of the horizon $r_+^q$ is not compatible
with the ansatz $r_+^q=MG+\k r_1$. If we take the ansatz in
the form $r^q=M+\sqrt{\k} r_1$, we get
\be r_1^2=-6-Mm_0\ .\ee
We see that, although this expression can be positive (for
$m_0<-{6\over M} $), it is physically rather unacceptable to have the
correction $r_1$ dependent on the arbitrary parameter $m_0$. This indicates
that the correction of the horizon is probably nonanalytic.

The components of EMT  for this solution are
\be T_{uu}=T_{vv}=-
{(r-{r_+})^4\over 48\p r^6{r_+}^2}(r^2+4r{r_+}+{r_+}^2+6{r_+}^2\log
{r\over l})\ ,\label{eq:flux} \ee
\be T_{uv}={r_+(r-r_+)^2(4r-3r_+)\over 48\p r^6} \ .\label{tt} \ee
The asymptotic value of energy density $T_{00}$ is
$-{1\over 48\p {r_+}^2}$ and does not vanish,
which means that the temperature of the radiation 
is probably not zero.
Further, the factor $(r-r_+)^4$ in (\ref{eq:flux})
ensures the
regular behaviour of all components of EMT in the free
falling observer frame at the horizon, in accordance with the condition
(\ref{frame}).
Note that in \cite{trivedi},
 where the effective action consists of the Polyakov-Liouville term only,
the choice of constants $C$ and $D$ which ensures the regular behaviour 
does not exist. This is a particular property of the  SSG model.

In order to calculate entropy, let us analyze the properties of the
Euclidean extension of the extreme \RN metric.
Near the horizon, $r_+^q,$ the Euclidean metric $ds^2=fe^{2\phi}d\tau
^2+{1\over f}dr^2$, where 
$\tau \in [0,2\p \bar \beta ]$ ($2\p \bar \beta =\bar T^{-1}$),
 does not posess the conical singularity and this means that the
temperature of the black hole is arbitrary \cite{HHR}. In order to find
the corresponding entropy, let us write
the Euclidean effective action
with the appropriate surface terms added (see Appendix of \cite{brm}):
\bea
\G _1&=&-{1\over 4G}{\Big(} \int _{\tilde M_\alpha }d^2x\sqrt g[
r^2R+2(\nabla r)^2+2U(r)]\ncr
&-&\kappa\int d^2x \sqrt g [ 2R(\psi -6\chi )+(\nabla \psi )^2 \ncr
 &-&12 (\nabla \psi
)(\nabla \chi) -12{\psi (\nabla r)^2\over r^2}-12R\log r]{\Big)}\ncr
&-&{1\over 2G}{\Big(}\int _{\partial \tilde {M_\alpha }}r^2kds -
\kappa
 \int _{\partial \tilde {M_\a}}(2\psi -12\chi -12\log r)kds\Big{)}\ .
\eea
After a simple calculation,
using the expression for
external curvature of the boundary of the manifold,
$k={f^{\prime}\over 2f}+\phi ^{\prime}\sqrt f,$
we get
\bea \G _1&=& {\p\bar \b\over G}\int _{r_+}^L dx [2rr^{\prime}(\sqrt f
e^{\phi})^{\prime} \sqrt f -fr^{\prime 2}-2U(r)]\ncr
&+& {\k \p \bar \b \over G}\int _{x_+}^L dx \Big( (\sqrt f
{e^{\phi}})^{\prime}(2\psi-12\chi-12 \log r)^{\prime}\ncr
&+&{1\over 2}f\psi ^{\prime 2}-6f\psi ^{\prime}\chi^{\prime}-6{\psi f
r^{\prime 2}\over r^2}\Big)\label{gama} \eea
From (\ref{gama})  we see that $\G _1$ is proportional to $\bar \b$,
which means
that the entropy is equal to zero. 
The zero entropy is connected to the fact that 
the proper distance between the horizon $r=r_+$ and any other point of
the manifold is infinite:
 extreme black hole has
zero entropy and arbitrary temperature. On the other hand it seems that
temperature of the scalar field gas is not arbitrary and this can be
a consequence of the fact that the heat capacity of the extreme black
hole is zero \cite{psstw}.

\section{Conclusions}

We calculated the quantum corrections for the \RN charged black hole. The
backreaction of the radiation is described by the SSG effective action
(\ref{eq:S}), in which the effects of nonlocality are expressed through
the auxilliary fields $\psi$ and $\chi$.

In the case of the nonextreme black hole the semiclassical procedure
for finding the correction is unambiguously defined and can be
performed similarly to the \Sch case. The \HH vacuum state 
 which describes the thermalized black
hole in equilibrium with radiation is defined by choosing the values of
constants $C$ and $D$. Here, also, the asymptotic value of the energy
density is negative, which is a characteristic of the SSG model as it
was discussed in \cite{brm,bf}.
The analysis of the extreme \RN case turns out to be more difficult.
First,  treatment along the lines of Zaslavskii gives some physically
unacceptable results. This is related with the fact that the extreme
black holes are topologically different from the nonextreme ones. 
In the framework of the Hawking approach,
the behaviour of the Euclidean extension
 shows that the temperature of the extreme \RN black hole is
arbitrary, while its entropy is zero. On the other hand, our analysis
of the EMT of radiation (in the SSG model) shows that there exists 
some value of temperature for which the energy-momentum tensor
is regular everywhere. But, at the same time, we see that the
correction of the position of the horizon is nonperturbative.
This is probably the consequence of the 
fact that the horizon of the
extreme black hole is not bifurcate. The coordinates
which are regular at the horizon are not of the Kruscal type
$U=e^{-\alpha u}$, $V=e^{\alpha v}$ \cite{c}, and the horizon consists
of the four unconnected branches. This might be the physical reason for
the impossibility of defining of the \HH vacuum 
by choosing special coordinates.

\sect{Appendix}

In this appendix we will analyze the general model where the conformal
term $\xi
f_i^2R$ is included in 4D action, i.e the action given by (\ref{eq:S4}).
 We want
to calculate the one-loop effective action taking the quantum
correction for the
matter field $f_i$ only.
In short, we will repeat the steps in \cite{mr4,mr3},
for this model. If we rescale the field of matter, $f_i\to e^{\Phi}f_i$
 we get

$$\li{ \G _m&=-{1\over 2}\sum_i\int d^2x
 \sqrt{-g}((\nabla f_i)^2+2f_i\nabla f_i\nabla
\Phi+f_i^2(\nabla \Phi)^2)\cr
&-{\xi\over 2}\sum_i\int d^2x\sqrt{-g}f_i^2(R+4\Box
\Phi-6(\nabla \Phi)^2+2e^{2\Phi})&(A.1)\cr}$$
Using the
complexification of the fields (the doubling trick) we get
$$ \li{\G _m&=-\sum _i\int d^2x\sqrt{-g}(\nabla f_i^*\nabla f_i-
\Box \Phi f_i^*f_i+(\nabla \Phi)^2f_i^*f_i)\cr
&-
\xi\sum _i\int d^2x\sqrt{-g}f_i^*f_i(R-6(\nabla \Phi )^2+
4\Box \Phi +2e^{2\Phi}) \ .&(A.2)
\cr}$$
Expanding the background metric around a flat metric, $g_{\m\nu}=\eta
_{\m\nu}+\g _{\m\nu}$ the action (A.2) becomes
$$\li{
\G _m&= \sum_i\int d^{2+\e }x f_i^*\Big[ \eta ^{\m\nu}\pa_{\m}\pa_{\nu} +
\overleftarrow{\pa_{\mu}
}\bar \gamma^{ \mu\nu}{\overrightarrow \pa}_{\nu}\cr
&+\sqrt{-g}(\Box \Phi-(\nabla \Phi)^2-
\xi (R-6(\nabla\Phi)^2+4\Box \Phi +2e^{2\Phi}))\Big] f_i\ ,&(A.3)\cr}$$
where $\bar \gamma^{ \mu\nu}=\gamma ^{\mu\nu}-{1\over 2}\gamma \eta
^{\mu \nu}.$
The vertices with two and zero spacetime derivatives we denote as A and
C (like as in \cite{mr4,mr3}). The corresponding diagrams are given by
$$ C=-{iN\p ^{{d\over 2}}\over (2\p)^2}\G(-{\e\over 2})\int
d^2x\sqrt{-g}(-\xi R-2\xi e^{2\Phi}-(1-6\xi )(\nabla \Phi)^2
+(1-4\xi)\Box \Phi)$$
$$ AC=-{iN\p^{{D\over 2}}\over (2\p )^2}\Bigl(\int d^2x \sqrt{-g}R{1\over
\Box}\Big[ -\xi R-2\xi e^{2\Phi}-(1-6\xi )(\nabla \Phi )^2\big] 
+(1-4\xi )\int d^2x
\sqrt{-g}R\Phi\Bigr)$$
$$ AA=-{iN\p ^{{D\over 2}}\over (2\p )^2}\G (1-{\e\over 2})B(2+{\e
\over 2},2+{\e\over 2})\int d^2x\sqrt {-g}\Big( R{1\over \Box}R+{4\over \e
(1+{\e\over 2})}R\Big)\ .$$
The one-loop correction to the effective action is given by \cite{mr4,mr3}
$$\G ^{(1)}={i\over 2}(C-AC-{1\over 2}AA)$$
Using (A.2) we get the effective action (\ref{eq:S2eff}).
Given this action we can investigate the vacuum solutions of the
corresponding equations of motion along the lines developed above and in
\cite{brm}. Unfortunately, this does not change our results essentialy.
The flux of radiation at large radius remains negative,
$T_{uu}={(r_+-r_-)(-5r_++r_-)\over 192 \p r_+^3}$, which was one point
where one might expect improvement. In the analysis of the extreme
solution one can see that the regular behavior of EMT 
exists only in the case
$\xi =0$, which is presented in detail.

\end{document}